\documentclass[twocolumn]{aastex63}

\newcommand{\xmm} {{\it XMM-Newton}}
\newcommand{\chandra} {{\it Chandra}}
\newcommand{\nustar} {{\it NuSTAR}}
\newcommand{\swift} {{\it Swift}}

\newcommand{\swiftxrt} {{\it Swift}/XRT}

\newcommand{\cmsq} {cm$^{-2}$}
\newcommand{\nh} {$N_{\rm{H}}$}
\newcommand{\lx} {$L_{\rm{X}}$}
\newcommand{\fx} {$F_{\rm{X}}$}

\newcommand{\ergs}{\mbox{\thinspace erg\thinspace s$^{-1}$}}
\newcommand{\ergcms}{\mbox{\thinspace erg\thinspace cm$^{-2}$\thinspace s$^{-1}$}}

\newcommand{\cntrt}{counts\,s$^{-1}$}

\newcommand{\msol} {$M_{\odot}$}

\shorttitle{Evolution of M51 ULX7}
\shortauthors{Brightman et al.}

\begin{document}

\title{Evolution of the spin, spectrum and super-orbital period of the ultraluminous X-ray pulsar M51 ULX7.}

\author{Murray Brightman}
\affiliation{Cahill Center for Astrophysics, California Institute of Technology, 1216 East California Boulevard, Pasadena, CA 91125, USA}

\author{Matteo Bachetti}
\affiliation{INAF-Osservatorio Astronomico di Cagliari, via della Scienza 5, I-09047 Selargius (CA), Italy}

\author{Hannah Earnshaw}
\affiliation{Cahill Center for Astrophysics, California Institute of Technology, 1216 East California Boulevard, Pasadena, CA 91125, USA}

\author{Felix F\"{u}rst}
\affiliation{Quasar Science Resources SL for ESA, European Space Astronomy Centre (ESAC), Science Operations Departement, 28692 Villanueva de la Cañada, Madrid, Spain}

\author{Marianne Heida}
\affiliation{European Southern Observatory, Garching, Germany}

\author{Gian Luca Israel}
\affiliation{INAF-Osservatorio Astronomico di Roma, via Frascati 33, 00078 Monteporzio Catone, Italy}

\author{Sean Pike}
\affiliation{Cahill Center for Astrophysics, California Institute of Technology, 1216 East California Boulevard, Pasadena, CA 91125, USA}

\author{Daniel Stern}
\affiliation{Jet Propulsion Laboratory, California Institute of Technology, Pasadena, CA 91109, USA}

\author{Dominic J Walton}
\affiliation{Institute of Astronomy, Madingley Road, Cambridge CB3 0HA, UK}

\email{murray@srl.caltech.edu}

\begin{abstract}
M51 ULX7 is among a small group of known ultraluminous X-ray pulsars (ULXP). The neutron star powering the source has a spin period of 2.8\,s, orbits its companion star with a period of 2 days, and a super-orbital period of 38 days is evident in its X-ray lightcurve. Here we present \nustar\ and \xmm\ data on the source from 2019 obtained when the source was near its peak brightness. We detect the pulsations, having spun up at a rate of 3$\pm0.5\times10^{-10}$ s\,s$^{-1}$ since they were previously detected in 2018. The data also provide the first high-quality broadband spectrum of the source. We find it to be very similar to that of other ULXPs, with two disk-like components, and a high energy tail. When combined with \xmm\ data obtained in 2018, we explore the evolution of the spectral components with super-orbital phase, finding that the luminosity of the hotter component drives the super-orbital flux modulation. The inclination the disk components appear to change with phase, which may support the idea that these super-orbital periods are caused by disk precession. We also reexamine the super-orbital period with 3 years of \swiftxrt\ monitoring, finding that the period is variable, increasing from 38.2$\pm0.5$ days in 2018--2019 to 44.2$\pm0.9$ days in 2020--2021, which rules out alternative explanations for the super-orbital period. 
 
\end{abstract}

\keywords{}

\section{Introduction}

M51 ULX7 was first detected as an X-ray source in the galaxies of M51 (NGC 5194/5) by the {\it Einstein} X-ray observatory \citep{palumbo85}. The source was observed with an X-ray luminosity of $\sim10^{39}$ \ergs, which the authors noted was brighter than any X-ray source in our own Galaxy, or M31, and that it exceeded the Eddington luminosity of a 1 \msol\ object. They suggested this X-ray source, and others like it, could be powered by a neutron star experiencing super-Eddington accretion in a non-spherically symmetric accretion flow geometry or could indicate the presence of a more massive black hole. These luminous X-ray sources would then come to be known as {\it ultraluminous X-ray sources}, or ULXs \citep[][see also recent reviews by \cite{kaaret17} and \cite{fabrika21}]{fabbiano89}. Thanks to the sensitivity and spatial resolution of \xmm\ and \chandra, we now know hundreds of ULXs \citep[e.g.][]{liu05,swartz11,walton11,earnshaw19,kovlakas20}.

While the assumption that ULXs were powered by black holes gained the most traction since their discovery by {\it Einstein}, the neutron star hypothesis was eventually proven, at least for a handful of sources. This began in 2014 with the detection of coherent pulsations from the ULX M82 X-2 by the \nustar\ observatory \citep{bachetti14}, determining the source to be powered by a neutron star, since black holes are incapable of producing such signals. This was followed by NGC~5907~ULX \citep{israel17a}, NGC~7793~P13 \citep{israel17,fuerst17}, NGC~300~ULX \citep{carpano18}, NGC~1313~X-2 \citep{sathyaprakash19}, and the subject of this paper, M51 ULX7 \citep{rodriguez20}. SMC X-3 \citep{tsygankov17}, Swift J0243.6+6124 \citep{wilsonhodge18} and RX~J0209.6-7427 \citep{vasilopoulos20b} also briefly became ULX pulsars, and another candidate pulsating ULX was also recently reported in NGC~7793 \citep{quintin21}.

From timing analysis of \xmm\ data, \cite{rodriguez20} determined the spin period of M51 ULX7 to be $2.8$\,s, and that the neutron star was in a 2-day orbit with a $>8$ \msol\ companion star, making it a high-mass X-ray binary system (HMXB). The long-term, secular spin up rate was also found to be $\sim10^{-9}$\,s\,s$^{-1}$. \cite{hu21} and \cite{vasilopoulos21} also found evidence for periodic dips in the \chandra\ X-ray light curve that are associated with the 2-day binary orbital period which they interpret as eclipses, implying that the orbit of the neutron star and its donor star is seen at high inclination.

The spin period is similar to the $\sim1$\,s spin periods of M82 X-2, NGC~5907~ULX, and NGC~7793~P13. All four sources also share the common characteristic of having periodic flux modulations in their long-term X-ray lightcurves, with the latter over a range of 60--80 days \citep{walton16b,fuerst18,brightman19}. The flux modulation from M51 ULX7 was found to have a period of 38 days from \swiftxrt\ monitoring \citep{vasilopoulos20,brightman20}. These periodic flux modulations, which in most cases are longer than the orbital period of the system, have been interpreted as precession of a large scale height disk, possibly caused by the Lense-Thirring effect \citep[e.g.][]{middleton18}. However, the origin of these super-orbital periods is still a matter of debate.

While the aforementioned ULXs have now been determined to be powered by a neutron star, it is still unknown what fraction, if any, are powered by black holes. From an X-ray spectral standpoint, the neutron star powered ULXs appear very similar to ones with unknown accretors, albeit among the hardest, implying that the vast majority could be powered by neutron stars \citep{pintore17,koliopanos17,walton18c,gurpide21}. The spectral shape consists of two disk-like components, a cooler one which may come from the outer regions of an accretion disk or the photosphere of an outflow \citep{qiu21}, and a hotter component, which may originate from the inner regions of the accretion disk \citep{walton18c}, an accretion curtain \citep{mushtukov17}, or Compton up-scattering \citep{titarchuk94}. A high energy tail is also seen when \nustar\ data are available, and appear to be associated with the pulsed component \citep{walton18c}.

While \nustar\ has observed M51 ULX7 on two previous occasions with lower energy coverage, the exposure time was either too short for a good quality spectrum \citep{earnshaw16}, or the source was caught at a low flux \citep{brightman18b}. In \nustar\ Cycle 5, we obtained joint \nustar\ and \xmm\ observations of M51 ULX7, timed to occur at the peak of the periodic flux modulation, with the aims of obtaining a high-quality broadband X-ray spectrum of ULX7, modeling its emission components, and tracking its pulsations. We present the results from these observations in this paper. We assume a distance of 8.58$\pm$0.10 Mpc to M51, derived from the tip of the red giant branch method \citep{mcquinn16}.

\section{X-ray data reduction}
\subsection{NuSTAR}
\label{sec_nustar}

\nustar\ \citep{harrison13} observed M51 from 2019-07-10 05:56:09 to 2019-07-14 01:26:09 with an exposure of 169 ks (obsID 60501023002). We used {\sc heasoft} v6.28, {\sc nustardas} v2.0.0 and {\sc caldb} v20201101 to analyze the data. We produced cleaned and calibrated events files using {\sc nupipeline} with the settings {\tt saacalc=3 saamode=OPTIMIZED tentacle=yes} to account for enhanced background during passages of the South Atlantic Anomaly (SAA), which reduced the exposure time to 162 ks. We used {\sc nuproducts} to produce spectral data, including source and background spectra, and response files. A circular region with a radius of 30\arcsec\ was used to extract the source spectra. A circular region with a radius of 100\arcsec\ was used to extract the background spectra, taking care to use extract the background from the same chip as the source. For timing analyses, we used the {\sc heasoft} tool {\tt barycorr} to apply a barycentric correction to the event times of arrival.

\subsection{XMM-Newton}
\label{sec_xmm}

\xmm\ \citep{jansen01} observed M51 from 2019-07-11 10:47:26 to 2019-07-12 08:09:24.000 with an exposure of 77 ks (obsID 0852030101). We used {\sc xmmsas} v18.0.0 to analyze the data. We first identify periods of high background by creating a lightcurve of the events in the 10--12 keV band, creating good time intervals where the rate was less than 0.1 \cntrt\ in this band, leaving 69 ks of data. Events were selected with {\tt PATTERN$\leq4$} for the pn and {\tt PATTERN$\leq12$} for the MOS. A circular region with a radius of 30\arcsec\ was used to extract the source spectrum. A circular region with a radius of 60\arcsec\ was used to extract the background spectra, on the same chip as the source and also in the galaxy in order to account for the soft diffuse emission the source is embedded in \citep[e.g.][]{earnshaw16}. Data from the pn and both MOS instruments were extracted in this way. For timing analyses, we used the {\sc xmmsas} tool {\tt barycen} to apply a barycentric correction to the event times of arrival.

\subsection{Swift}
\label{sec_swift}

\begin{figure}
\begin{center}
\includegraphics[trim=20 20 20 20, width=85mm]{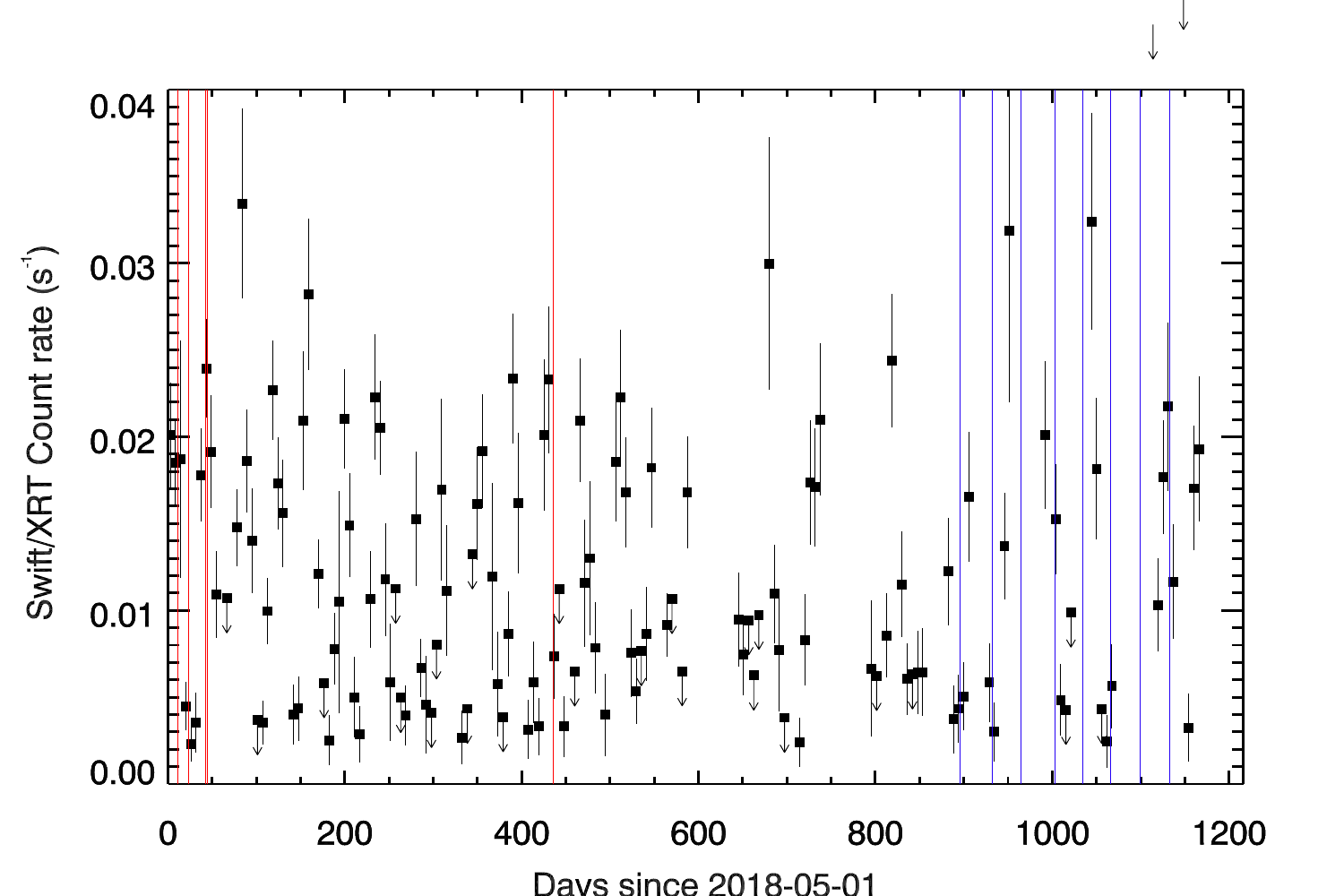}
\end{center}
\caption{\swiftxrt\ lightcurve of ULX7 over the period May 2018 -- May 2021 (black data points). Downward pointing arrows show 2.5$\sigma$ upper limits. \xmm\ and \chandra\ observation times are marked with red and blue lines respectively.}
\label{fig_ltcrv}
\end{figure}

We used the online tool provided by the University of Leicester\footnote{https://www.swift.ac.uk/user\_objects/} \citep{evans07,evans09} to extract the \swiftxrt\ \citep{burrows05} lightcurve of ULX7. All products from this tool are fully calibrated and corrected for effects such as pile-up and the bad columns on the CCD. We selected observations with target IDs 11417, 30083 and 32017, and binned the lightcurve in time, with the maximum bin size of 500 ks (5.79 days) and a minimum detection of 2.5$\sigma$. The lightcurve is plotted in Figure \ref{fig_ltcrv}, with the time of the 2019 \nustar\ and \xmm\ observation marked, as well as the 2018 \xmm\ observations. 

\section{Pulsation Analysis}
\label{sec_timing}

We searched for pulsations using the fast $Z^2_n$ search implemented in {\sc hendrics} tool {\tt HENzsearch} \citep{bachetti15}. This tool folds the data along a grid of frequencies and frequency derivatives and calculates the $Z^2_n$ statistic starting from the folded profiles \citep[see][]{huppenkothen19,bachetti21}. We used $n=1$ (Rayleigh test) and folded profiles of 16 bins (adequate for using the $Z^2_1$ search with the binned approximation). 

We found a clear peak in the $f$-$\dot{f}$ plane in the \xmm\ data, however, no such signal was found in the lower-count \nustar\ data. We noted the best solution from {\tt HENzsearch} and used {\tt HENphaseogram} to calculate the pulse time-of-arrival (TOA) in 16 intervals during the observation. Following this the graphical tool {\tt pintk} in {\sc pint} \citep{luo21} was used to fit the TOAs and get a timing solution and its uncertainty. We find $f = 0.358842(5)$\,Hz ($P=2.78674$\,s), and $\dot{f}=-1.19(6)\times10^{-8}$. We used the solution from {\tt pint} to phase tag the events using {\tt HENphasetag}.

We did not find evidence for the orbit of the neutron star and its companion when searching for a second derivative in the pulse frequency. This is likely due to too few counts and only a fraction of the orbit being covered by the \xmm\ observation. Therefore we could not make a correction for the orbit in the determination of the spin period. This means that the observed first derivative in the pulse frequency is the sum of both the secular spin up of the neutron star and the orbital motion. \cite{rodriguez20} calculated the maximum delay/advance introduced when not correcting for the orbital motion, which was of the order of 1\,ms. In the following, when inferring the secular spin up of the pulsar, we assume this 1\,ms value to be the absolute uncertainty of the period.

\begin{figure}
\begin{center}
\includegraphics[trim=20 20 20 20, width=85mm]{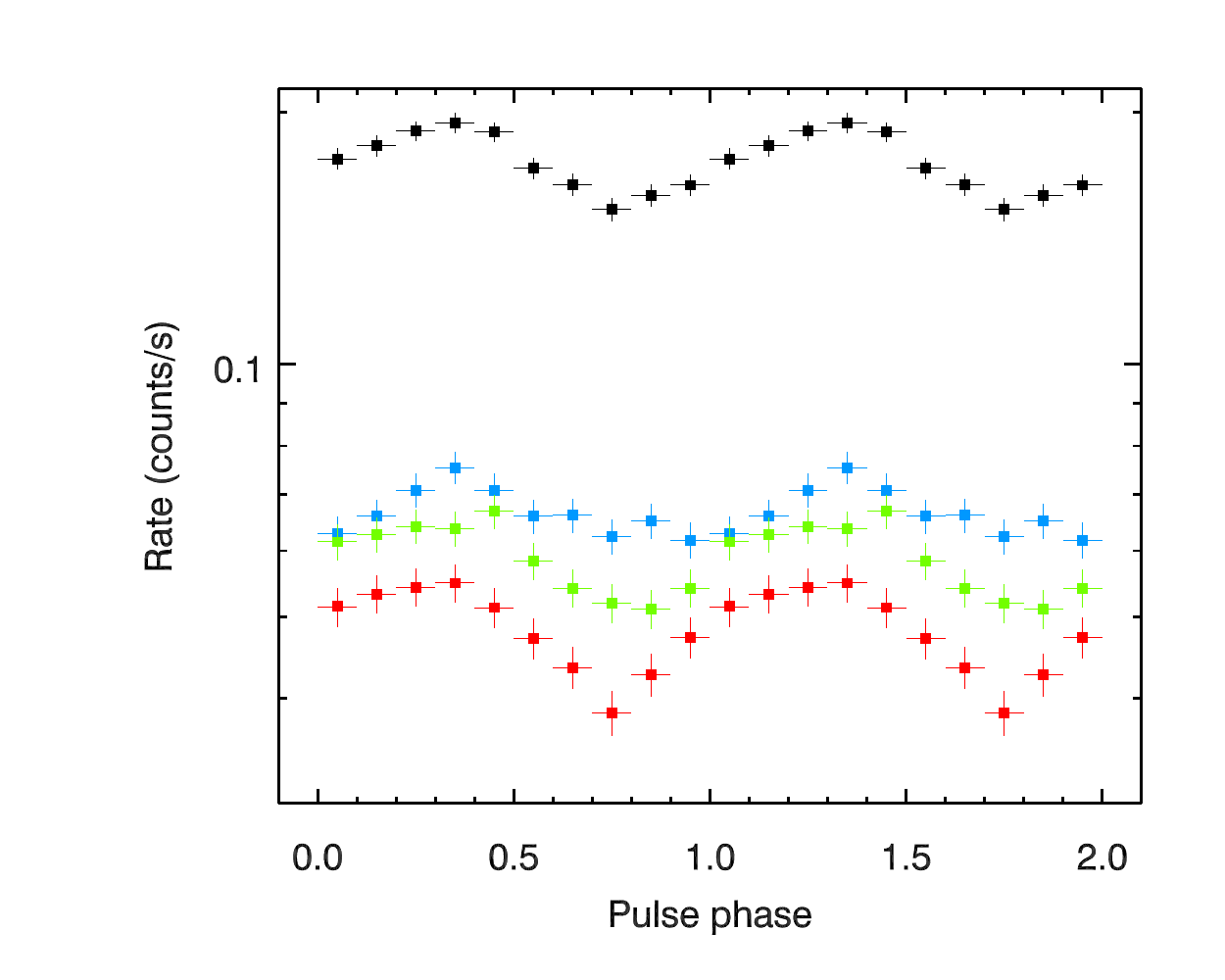}
\end{center}
\caption{The pulse profile of ULX7 in energy bands of 0.5--10 keV (black), 0.5--0.1 keV (blue), 1-2 keV (green) and 2--10 keV (red).}
\label{fig_pulse}
\end{figure}

\begin{figure}
\begin{center}
\includegraphics[trim=20 20 20 20, width=85mm]{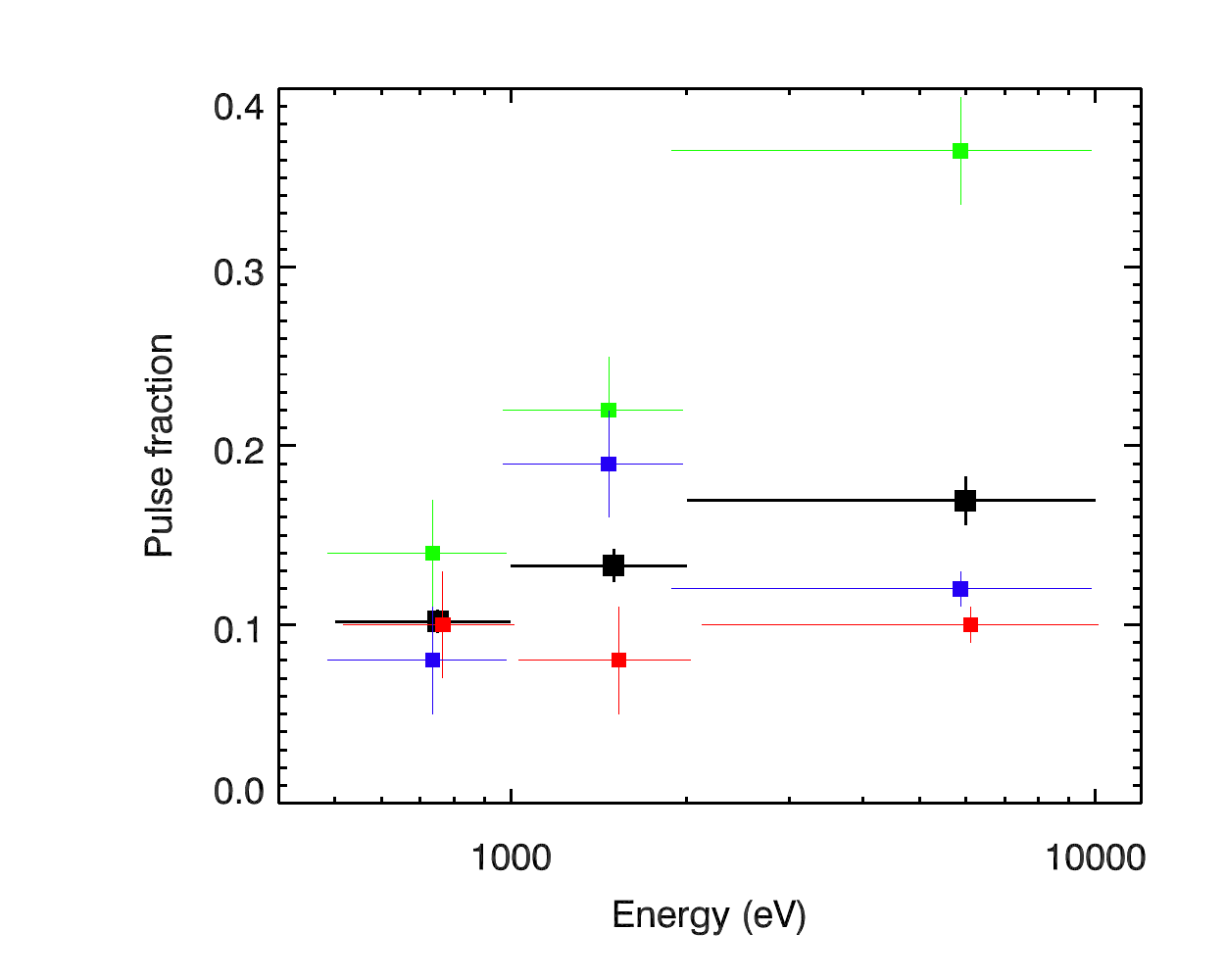}
\end{center}
\caption{The pulsed fraction of ULX7 versus energy from the 2019 \xmm\ data (black), and in comparison to the 2018 data from \cite{rodriguez20} (green, red, and blue for each obsID, slightly offset in energy for clarity). The pulsed fraction is highly variable and generally increases with energy.}
\label{fig_pf}
\end{figure}

Using the above timing solution and the phase-tagged events files, we create pulse profiles by binning the lightcurve in 16 equally sized phase bins, and energy bins of 0.5--0.1 keV, 1-2 keV and 2--10 keV. We plot these in Figure \ref{fig_pulse}. Using these, we also calculate the pulse fraction in each bin, defined as the amplitude of the pulse divided by the mean count rate. These are plotted in Figure \ref{fig_pf}. Also shown are the pulse fractions as determined by \cite{rodriguez20} from the 2018 \xmm\ observing campaign. We derive an upper limit of 74\% to the pulse fraction in the 10--20 keV band from the \nustar\ data by searching 170 Fourier frequencies,  and assuming $\dot{P}=\pm2\times\dot{P}_{max}$.

\section{Spectral analysis}
\label{sec_spectral}

We begin analyzing the 2019 \nustar\ Cycle 5 \nustar\ and \xmm\ spectral data on ULX7 by grouping the spectra with a minimum of 1 count per bin. We load the FPMA, FPMB, pn, MOS1 and MOS2 source spectra into {\sc xspec}, subtracting the background, and consider energies 3--20 keV for the \nustar\ data (the background dominates the source above 20 keV so we do not consider these data), and 0.2--10 keV for the \xmm\ data. We use the C-statistic, suitable for the low number of counts per bin here and we use a constant term to take account of cross-calibration uncertainties between instruments, which are typical of those found in \citep{madsen15}.

\cite{rodriguez20} carried out fits to the series of high-quality \xmm\ spectra they obtained in 2018 when the pulsations were detected. Their best-fit model consisted of an absorbed disk black body model plus a hotter blackbody model ({\tt tbabs*ztbabs*(diskbb+bbodyrad})), which we fit here. The {\tt tbabs} model accounts for absorption in our Galaxy, fixed at $3.19\times10^{20}$ \cmsq\ \citep{HI4PI16} and the {\tt ztbabs} accounts for absorption at the redshift of M51, $z=0.002$, left as a free parameter. This model resulted in $C=4027.81$ with 4286 d.o.f. 

\begin{figure}
\begin{center}
\includegraphics[trim=20 20 20 20, width=85mm]{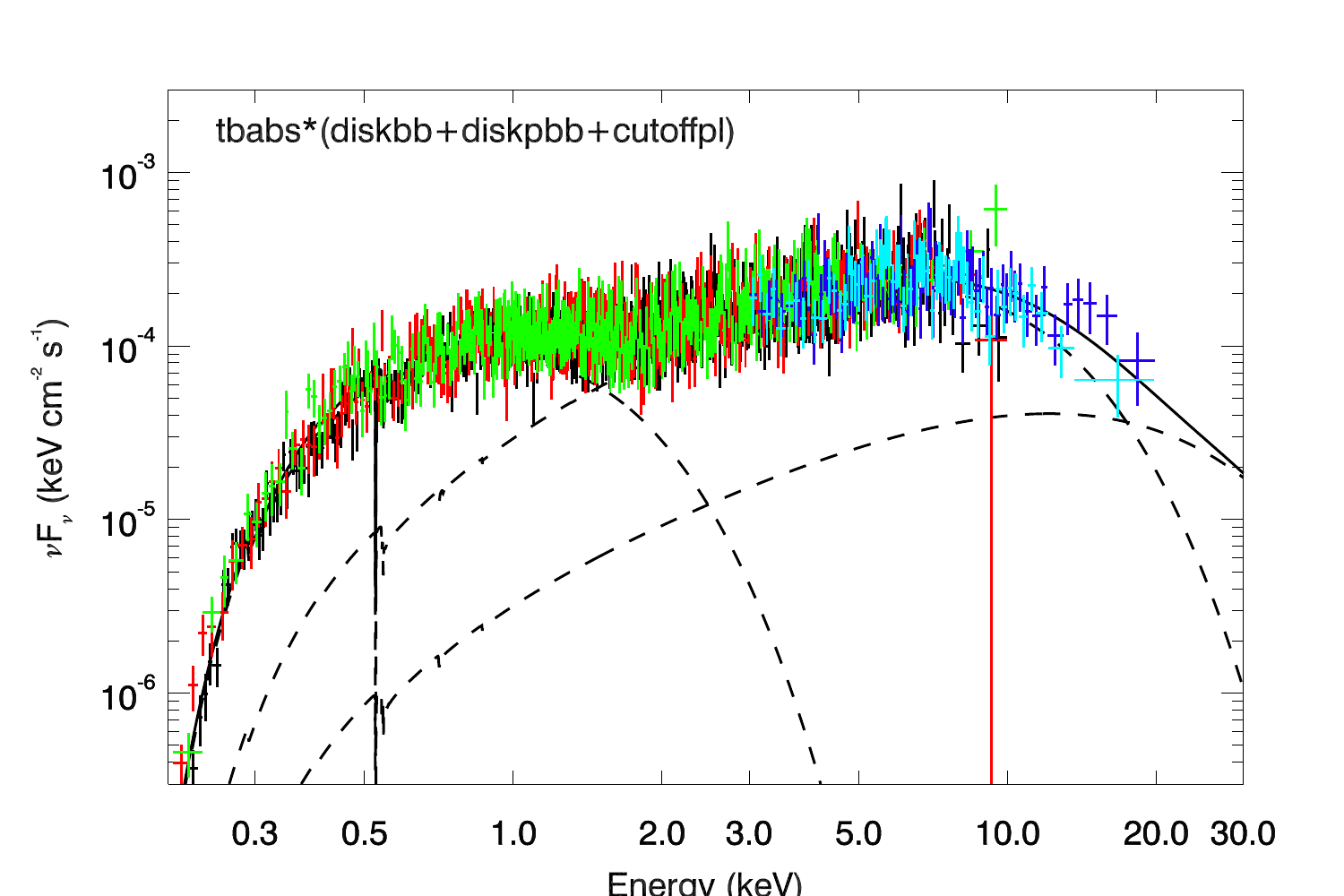}
\includegraphics[trim=20 20 20 0, width=85mm]{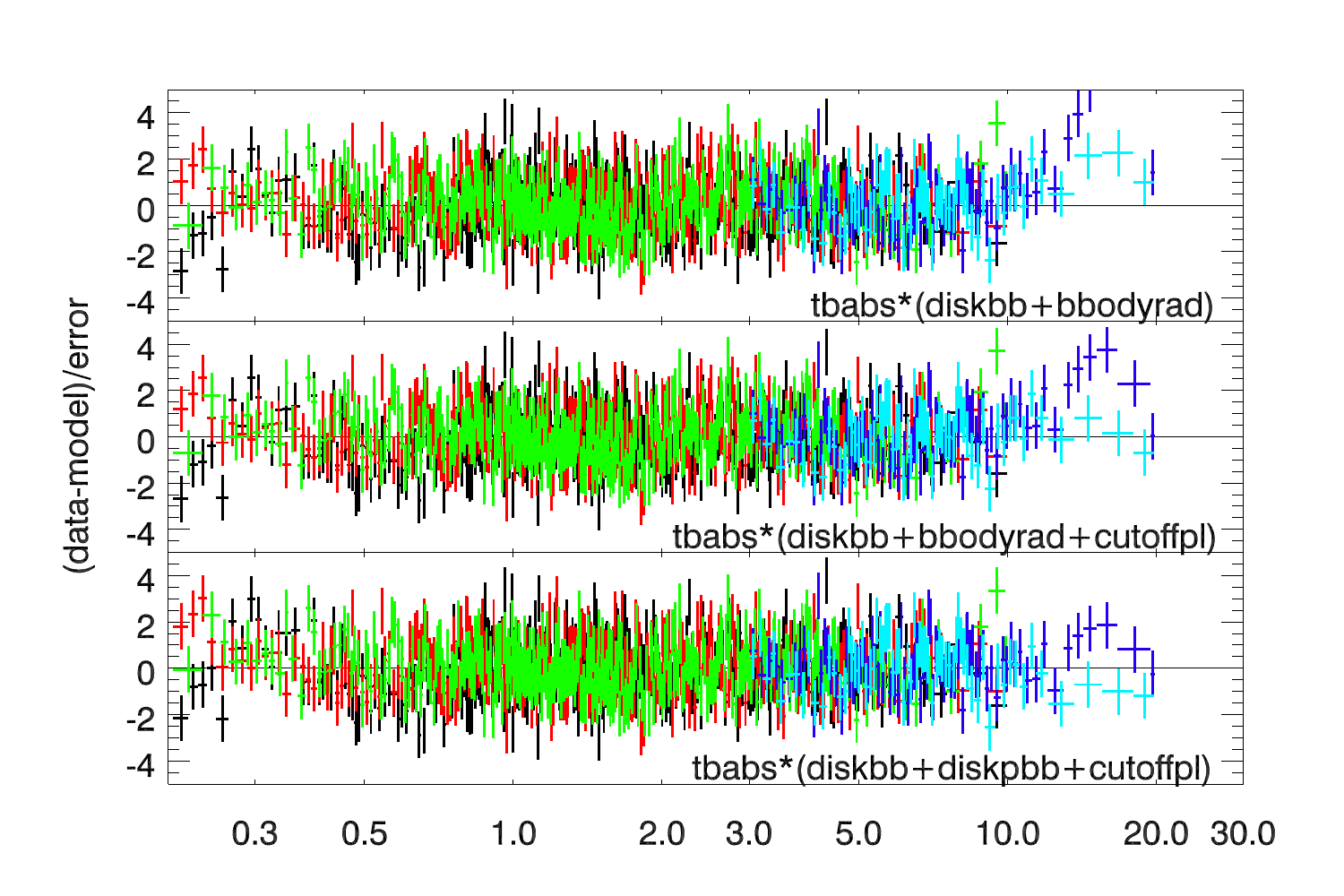}
\end{center}
\caption{Top - \xmm\ (black, red, and green), and \nustar\ (blue) spectra of M51 ULX7. The spectra can be modeled by the typical ULX spectrum which consists of two disk-like components, one at lower energy, the other at higher energy. A third component seen at the highest energies is the pulsed component, isolated using phase-resolved spectroscopy, defined in {\sc xspec} as {\tt tbabs*(diskbb+diskpbb+cutoffpl)}. Bottom - Data to model residuals for the models described in Section \ref{sec_spectral}.}
\label{fig_spec_ulx7}
\end{figure}

However, there appears to be an excess at energies above 10 keV when fitting with this model to the 2019 \nustar+\xmm\ data (Figure \ref{fig_spec_ulx7}).  \cite{rodriguez20} noted that in \cite{walton18c} a third component, a powerlaw with a high energy cut off ({\tt cutoffpl}) is used to model this component which is attributable to the accretion column. This pulsed component was isolated in the ULX pulsars M82 X-2, NGC 7793 P13 and NGC 5907 ULX by extracting spectra from the brightest and the faintest quarters ($\Delta\phi_{\rm pulse}=0.25$) of the pulse cycle and subtracting the latter from the former \citep[i.e., pulse on $-$ pulse off,][]{brightman16,walton18a,walton18c}. 

\begin{figure}
\begin{center}
\includegraphics[trim=20 20 20 20, width=85mm]{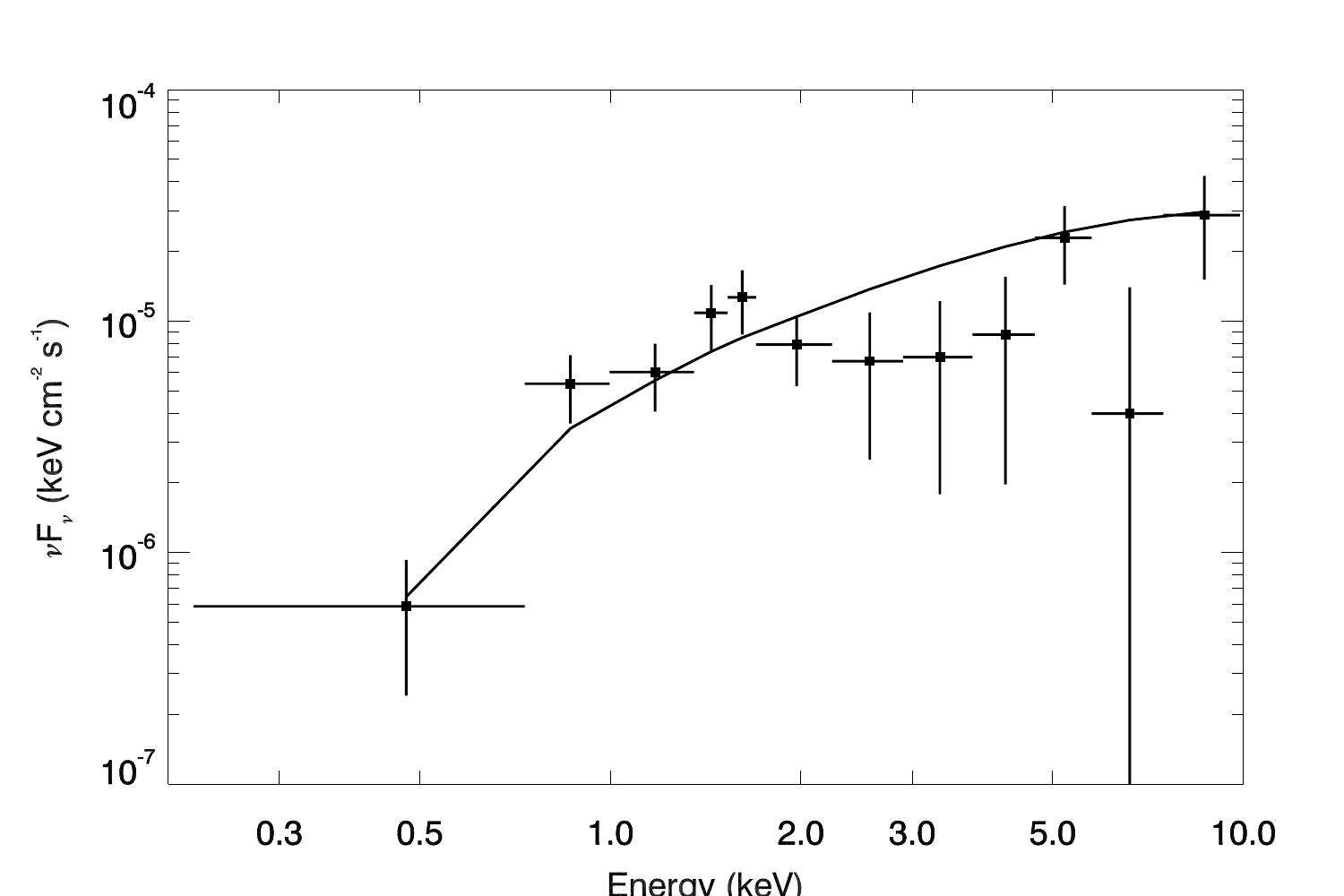}
\end{center}
\caption{Pulsed spectrum of ULX7 produced by the pulse on $-$ pulse off analysis of the \xmm\ data.}
\label{fig_pulsed_spec_ulx7}
\end{figure}

We do the same pulse on $-$ pulse off analysis here for the \xmm\ data on ULX7 using the timing solution of the pulses found in Section \ref{sec_timing}, and fit it with {\tt tbabs*ztbabs*cutoffpl}, where the absorption components have been fixed to the values from the time averaged spectrum. We cannot constrain $E_{\rm cut}$ due to the lack of high energy coverage, so we fix $E_{\rm cut}=8.1$, the average values from \cite{walton18c}. We can constrain $\Gamma=0.8\pm0.3$ with $C=881.89$ with 946 d.o.f., which is consistent with the average value of 0.5 from \cite{walton18c}. We plot this spectrum in Figure \ref{fig_pulsed_spec_ulx7}.

We then add this component to the time averaged spectral fit, keeping $\Gamma$ and $E_{\rm cut}$ fixed, but leaving the normalization free. This results in $C=3954.59$ with 4285 d.o.f., presenting an improvement of the overall time-averaged fit. 

We note that if a {\tt diskpbb} model were used for the hottest disk component instead of {\tt bbodyrad}, consistent with the models used in \cite{walton18c}, we get a better fit of $C=3951.17$ with 4284 d.o.f. Therefore, we present this {\tt diskbb+diskpbb+cutoffpl} model as our best fit.

\begin{table*}
\centering
\caption{Results from the X-ray spectral modeling of joint \xmm\ and \nustar\ data on ULX7}
\label{tab_specpar_ulx7}
\begin{center}
\begin{tabular}{c c c c }
\hline
Parameter 					& \multicolumn{3}{c}{Model}				 \\
							& {\tt diskbb+bbodyrad} 	& {\tt diskbb+bbodyrad+cutoffpl} & {\tt diskbb+diskpbb+cutoffpl}\\
\nh/$10^{20}$\cmsq\				& 1.3$\pm$0.7			& 3.3$\pm$0.7 			& 3.9$^{+1.3}_{-0.6}$ \\
$T_{\rm in, cool}$/keV			& 0.46$\pm0.02$		& 0.36$\pm0.02$ 		& 0.33$\pm0.03$	\\
$N_{\rm cool}/10^{-4}$ $^a$		& 4050$^{+500}_{-480}$ 	& 950$^{+350}_{-250}$ 	& 13500$^{+6200}_{-1800}$\\
$T_{\rm hot}$/keV				& 1.64$\pm0.05$		& 1.39$\pm0.09$		& 2.26$^{+0.54}_{-0.17}$ \\
$p$							& -					& -					& 0.99$^{+u}_{-0.25}$ \\
$N_{\rm hot}/10^{-4}$ $^a$		& 75.6$^{+10.6}_{-9.5}$ 	& 71.2$^{+17}_{-15}$ 	& 17.2$^{+17}_{-12}$\\
\fx/$10^{-13}$ \ergcms\	$^b$		& 8.3$\pm0.3$			& 8.2$\pm0.3$			& 7.7$\pm0.2$\\
\lx/$10^{39}$ \ergs\	$^c$			& 7.7$\pm0.4$			& 7.6$\pm0.4$ 			& 7.8$\pm0.4$\\
$C_{\rm pn}$ 					& 1.00$^{+0.07}_{-0.06}$	& 0.99$\pm0.07$		& 0.99$^{+0.07}_{-0.06}$\\
$C_{\rm MOS1}$ 				& 1.03$^{+0.08}_{-0.07}$ 	& 1.04$^{+0.08}_{-0.07}$	& 1.03$^{+0.08}_{-0.07}$ 	\\
$C_{\rm MOS2}$ 				& 1.05$^{+0.08}_{-0.07}$ 	& 1.07$^{+0.08}_{-0.07}$	& 1.05$^{+0.08}_{-0.07}$ 	\\
$C_{\rm FPMB}$ 				& 1.00$\pm0.08$ 		& 1.02$^{+0.09}_{-0.08}$	& 1.00$\pm0.08$ 		\\
$C$-statistic/d.o.f.				& 4027.81/4286		& 3954.59/4285		& 3951.17/4284 \\		
\hline
\end{tabular}
\tablecomments{$+u$ indicates a parameter has hit its upper bound in the fit. $^a$ In units of  ($R_{\rm in}$/D$_{10})^2$cos$\theta$ $^b$ observed in the 0.3--10 keV band. $^c$ intrinsic, corrected for absorption in the 0.2--20 keV band assuming a distance of 8.58 Mpc.}

\end{center}
\end{table*}

\section{Evolution of the spectral parameters}
\label{sec_ulx7_specevol}

With the newly determined best-fit model found in Section \ref{sec_spectral}, we proceed to apply it to the high-quality data from \xmm\ in 2018 in order to explore the evolution of the X-ray spectral parameters with luminosity and super-orbital phase. While these data were presented by \cite{rodriguez20}, they used a slightly different model, so we cannot strictly compare their results to ours. We note that these models are purely phenomenological, and any physical interpretations should have this added caveat.

We reduce the \xmm\ obsIDs 0824450901, 0830191401, 0830191501, 0830191601 in the same way as described in Section \ref{sec_xmm}. We then fit the spectra with the {\tt diskbb+diskpbb+cutoffpl} model described above. As found by \cite{rodriguez20}, ULX7 was seen in a low flux state in obsID 0830191401, and only the low temperature {\tt diskbb} can be seen. We remove the {\tt cutoffpl} model from the fit since it clearly over fits the data, and we fix the temperature and $p$ parameter of the {\tt diskpbb} model to typical values in order to place an upper limit on the normalization and flux of this component.

For each epoch, we calculate the total intrinsic luminosity and the luminosity of the two disk components separately, determined using the model component {\tt cflux} and calculated in the 0.3--10 keV range. To get the super-orbital phase, we divide the number of days since 2018-05-01 by the period of 38.3 days, and subtract off the nearest integer. We also adjust the phase so that phase=0 corresponds to the peak of the super-orbital modulation. In \cite{brightman20}, we showed that the super-orbital period of 38.3 days was consistent across the 2018--2019 epoch we study here.

\begin{figure}
\begin{center}
\includegraphics[trim=20 10 20 20, width=85mm]{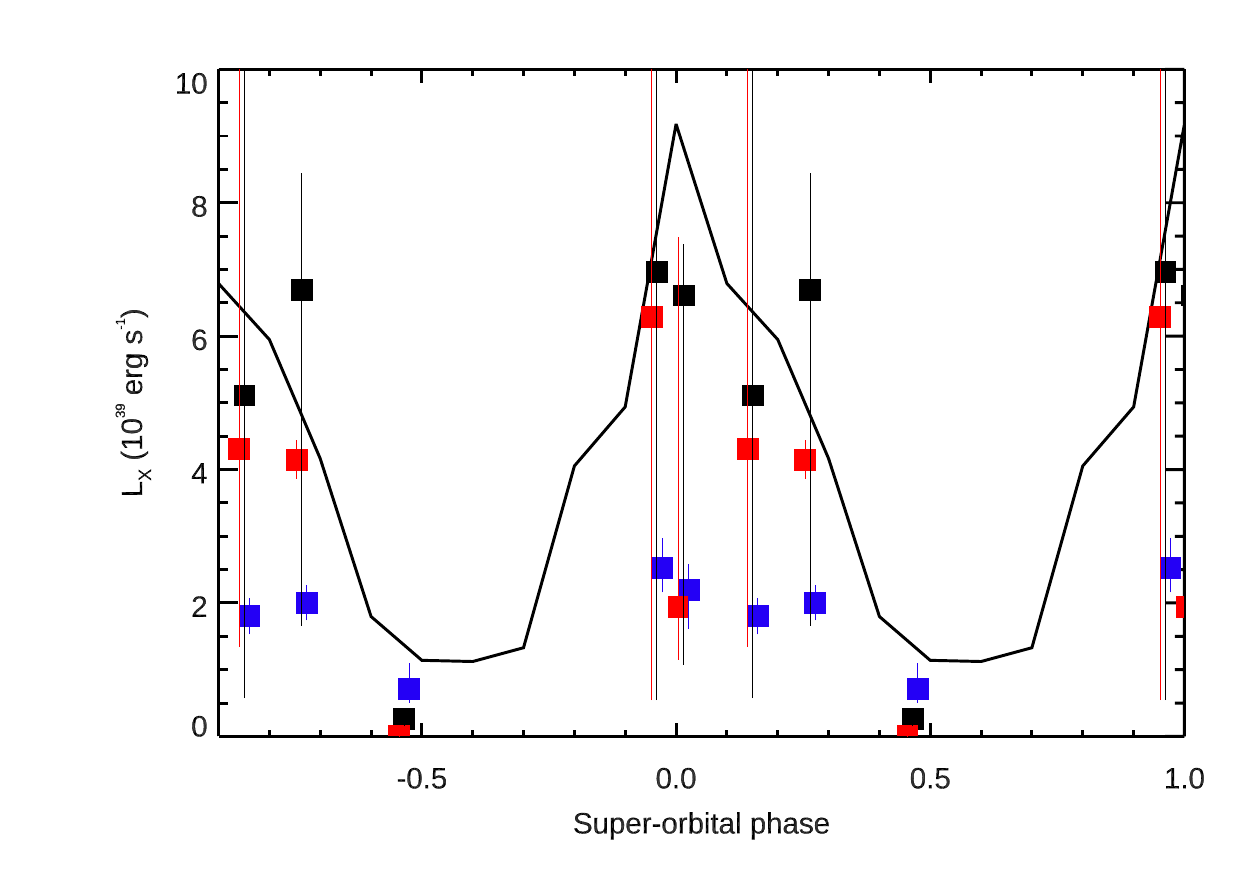}
\includegraphics[trim=20 20 20 20, width=85mm]{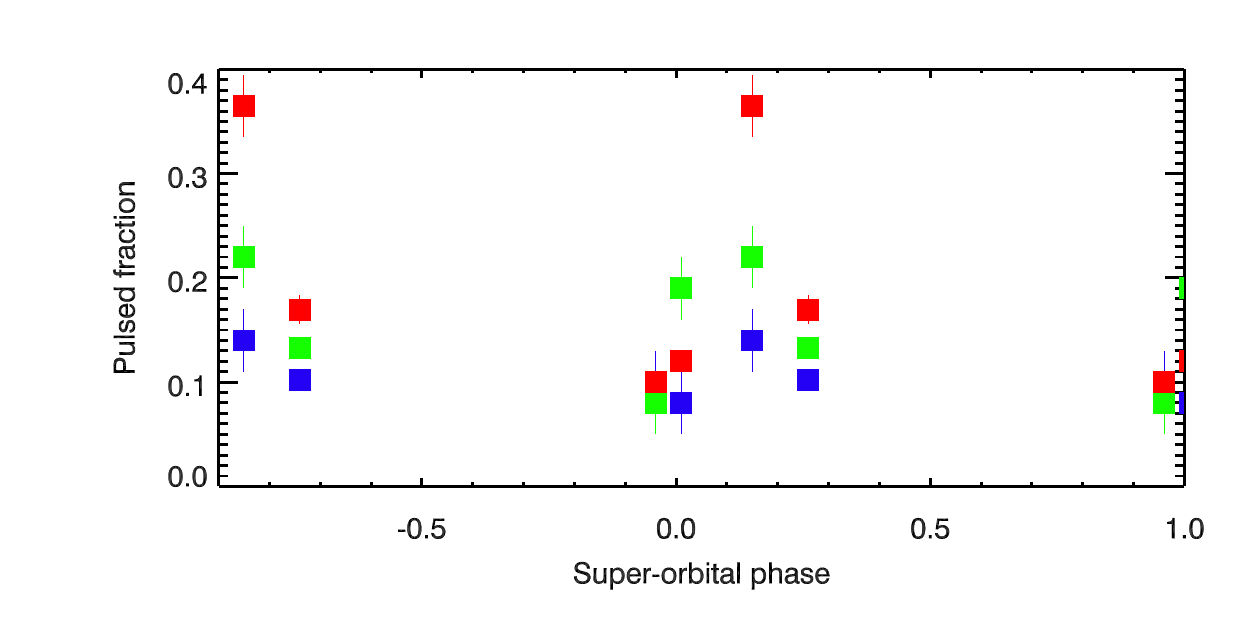}
\end{center}
\caption{Top - The total luminosity (black squares) and luminosity of the cool (blue squares) and hot (red squares) disk components of ULX7 as a function of its super-orbital phase from \nustar\ and \xmm\ observations in 2018--2019 (offset in phase for clarity). The average luminosity as a function of phase from \swiftxrt\ observations is shown with a black line. Bottom - The pulsed fraction in the 0.5--1 keV band (blue), 1--2 keV band, and 2--10 keV band (red) as a function of super-orbital phase with data from \cite{rodriguez20}.}
\label{fig_phase_flux}
\end{figure}

First we plot the total intrinsic luminosity, and the luminosity of the two disk components as a function of super-orbital phase, and compare these to the average total luminosity as seen from the \swiftxrt\ monitoring.  Figure \ref{fig_phase_flux} shows that the total luminosity measured by \xmm\ agrees in general with that of the average luminosity seen by \swiftxrt. In terms of the two disk components, the cool component's luminosity remains relatively constant, whereas the hot component's luminosity varies more, albeit with large uncertainties.

We also plot the pulsed fraction derived in Section \ref{sec_timing} and in \cite{rodriguez20}, and how it varies with super-orbital phase in Figure \ref{fig_phase_flux}. We see a similar picture here, where the pulsed fraction in the low energy band varies relatively little over the super-orbital period, but the higher energy pulsed fraction varies strongly. This is likely driven by the variation in the non-pulsed, hot disk component.

\begin{figure}
\begin{center}
\includegraphics[trim=20 40 20 20, width=85mm]{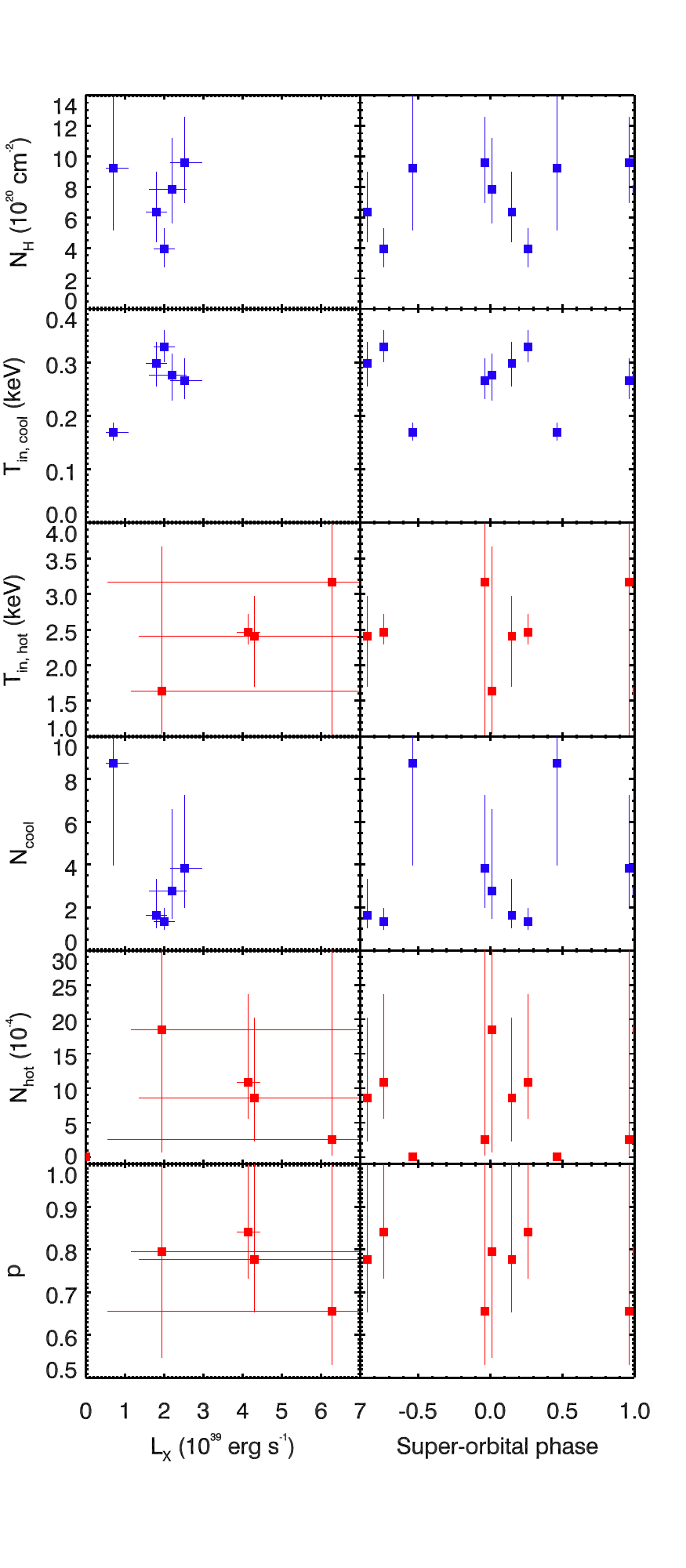}
\end{center}
\caption{Spectral parameters of the {\tt diskbb} (cool component, blue data points) and {\tt diskpbb} (hot component, red data points) models fitted to the 2019 \nustar\ and \xmm\ data, and 2018 \xmm\ data, plotted against the \lx\ for each individual component (left, 0.3--10 keV unabsorbed) and super-orbital phase (right, two cycles plotted for clarity). Uncertainties are plotted at the 90\% confidence level.}
\label{fig_specpar_ulx7}
\end{figure}

We then proceed to plot the spectral parameters of the disk components against the intrinsic luminosity and super-orbital phase in Figure \ref{fig_specpar_ulx7}. We do not see any statistically significant correlations between the spectral parameters and \lx\ for either the cool {\tt diskbb} component or hot {\tt diskbb} component. We do see that \nh, $T_{\rm in, cool}$, $N_{\rm cool}$ and $N_{\rm hot}$ appear to vary with the phase of the super-orbital period, however we likely do not have enough data to claim a significant relationship. If the dependence is real, then since the normalization of the disk components depends on the orientation of the disk, this may imply that the inclination of the disks os changing with super-orbital phase, or in other words, the disks are precessing.

\begin{figure}
\begin{center}
\includegraphics[trim=20 20 20 20, width=85mm]{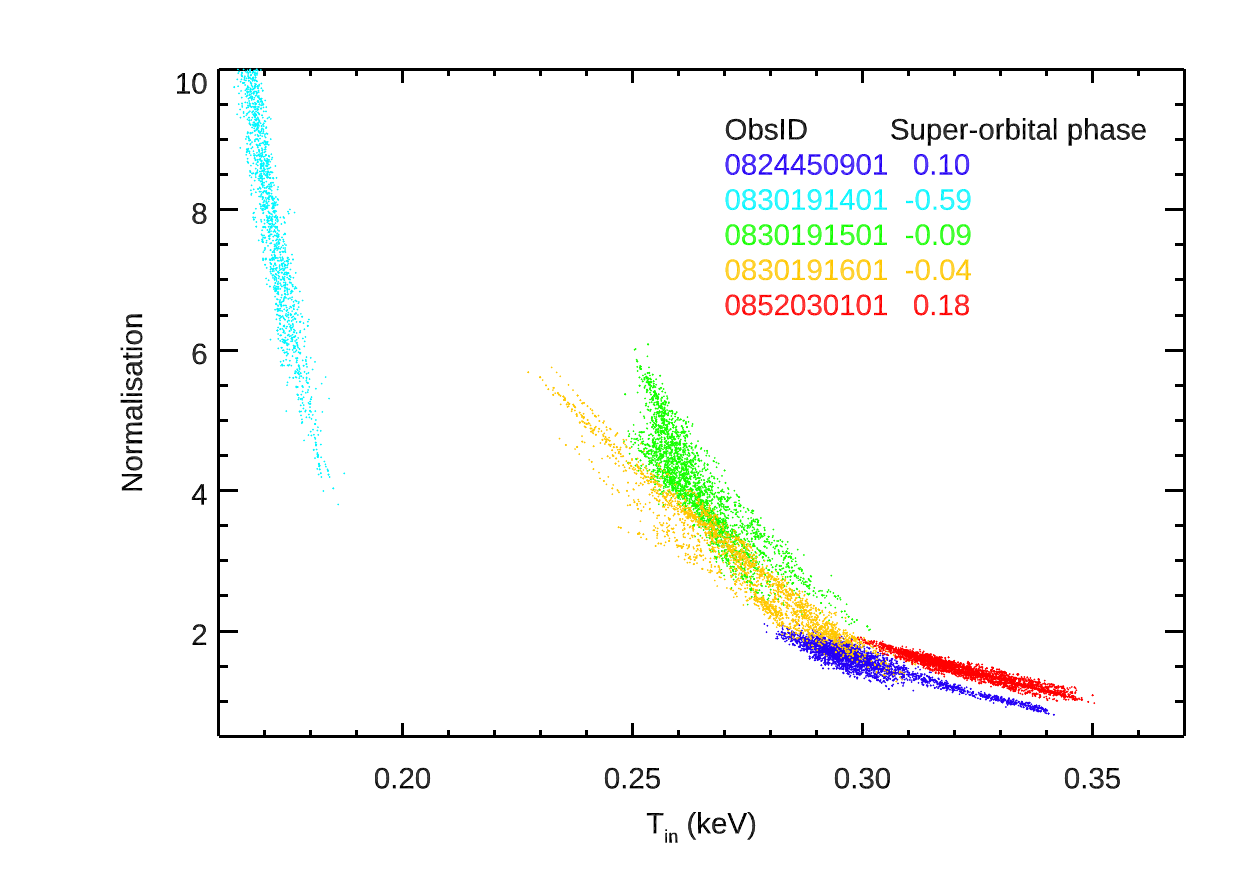}
\end{center}
\caption{Results of a 10000-step MCMC analysis for the $T_{\rm in, cool}$ and $N_{\rm cool}$ parameters from each \xmm\ observation. While a degeneracy can be seen between the two parameters, the results map out regions which are almost mutually exclusive, implying that the degeneracy is not the cause of the possible dependence of these parameters on super-orbital phase.}
\label{fig_specpar_con}
\end{figure}

However, since the disk temperature and normalization parameters are degenerate with each other, it is possible that these variations are driven by the degeneracy instead. We investigate this possibility by using Markov-Chain Monte-Carlo methods to map out the $T_{\rm in, cool}$ and $N_{\rm cool}$ parameter space. In {\sc xspec} we use the Goodman-Weare algorithm with 8 walkers and a total length of 10000 steps with a burn-in phase of 5000 steps. Figure \ref{fig_specpar_con} shows the results of this. While a degeneracy can be seen between the two parameters, the results map out regions which are almost mutually exclusive, implying that the degeneracy is not the cause of the possible dependence of these parameters on super-orbital phase.

\section{Evolution of the super-orbital period}
\label{sec_ulx7_superevol}

\begin{figure*}
\begin{center}
\includegraphics[trim=20 0 20 20, width=180mm]{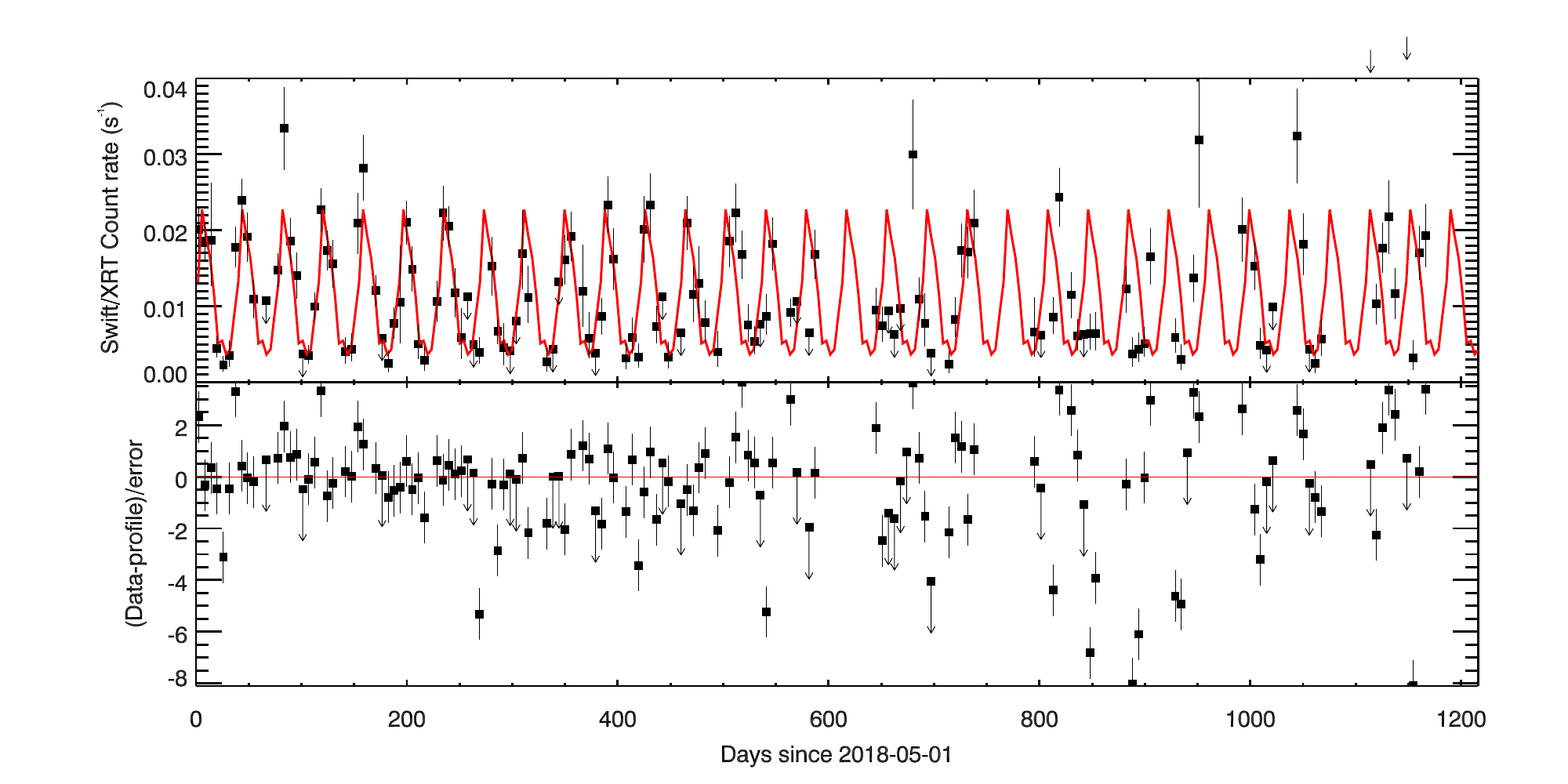}
\end{center}
\caption{Top - \swiftxrt\ lightcurve of ULX7 over 3-years 2018--2021 (black data points) with the average profile of the 38.2-day super-orbital period seen in the first 500 days overplotted (red line). Bottom - Residuals of the data to the profile. The data clearly follow the average profile for the first 500 days, but deviate thereafter, which we find is in part due to a change of the super-orbital period.}
\label{fig_ltcrv2}
\end{figure*}

In \cite{brightman20}, we showed that ULX7 exhibited a super-orbital period of 38 days over 500 days ($\sim$1.5 years) from May 2018 \citep[see also][]{vasilopoulos20}. However, when we applied our analysis to the most recent, full $\sim3$-year data set, the signal appears much weaker, with a peak in the $L$ statistic at 34.1 days and a peak in the Lomb-Scargle periodogram at 39.0 days. This implies that after 500 days, the super-orbital period disappears, or changes phase and or period. Figure \ref{fig_ltcrv2} shows the full 3-year lightcurve with the average profile of the flux modulations from the first 500 days overplotted. The deviation from the profile after 500 days is clear in the residuals, and exhibits both positive and negative deviations therefore cannot be explained by the anomalous low flux states seen by \cite{vasilopoulos20}.

We investigate further by splitting the lightcurve into smaller sections in time to determine how the super-orbital period evolves. As in \cite{brightman20}, we use epoch folding on the $L$ statistic \citep{davies90}, using 10 phase bins. This time however, since we know the approximate period, we search over a narrower range of 30--50 days in 200 equally spaced bins for the epoch folding. We find that a 300-day section is sufficient to recover the 38-day period at the beginning of the lightcurve, and then we progress through the lightcurve with steps of 30 days. 

\begin{figure*}
\begin{center}
\includegraphics[trim=20 20 20 20, width=180mm]{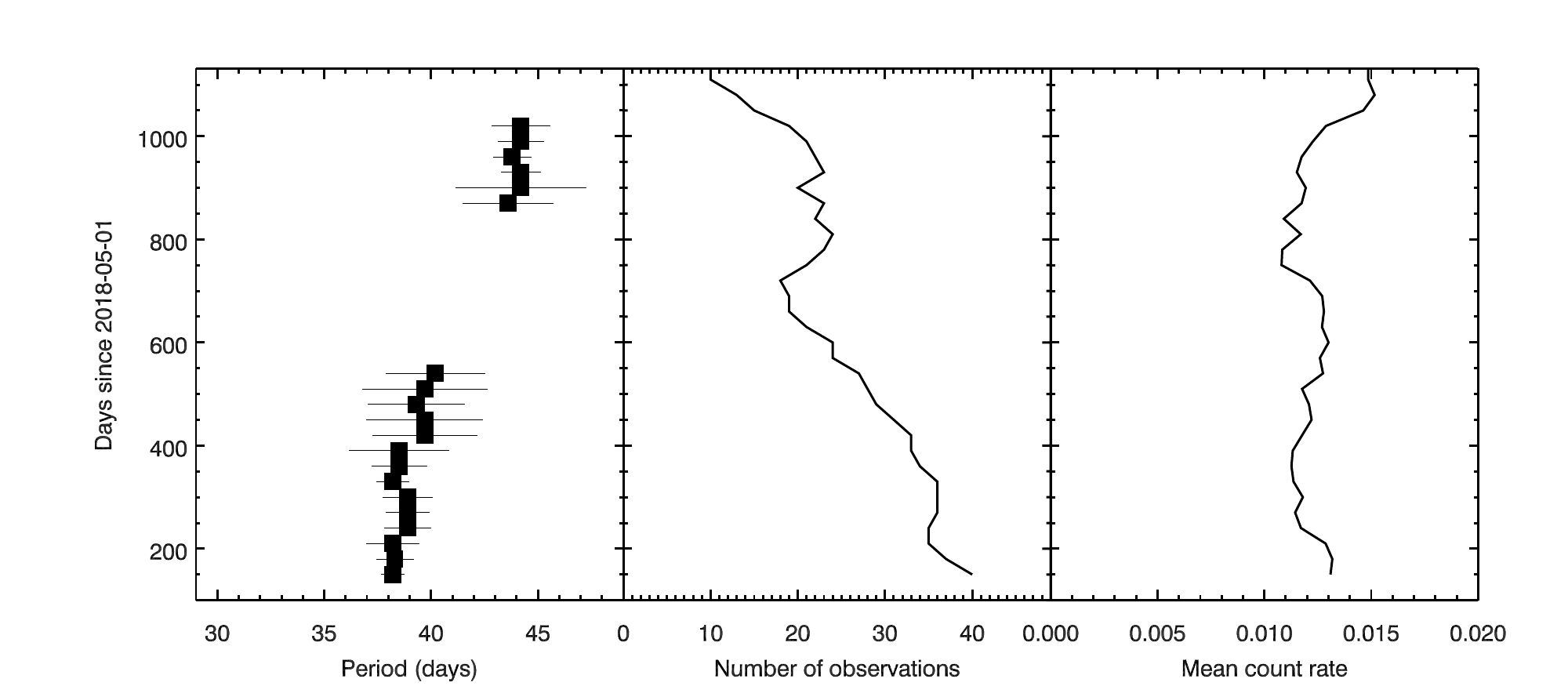}
\end{center}
\caption{Left - The evolution of the super-orbital period, where each data point represents a 3$\sigma$ detection for a 300-day time bin. The detection significance is determined by simulations. Error bars are 1$\sigma$, and are also determined by simulations. The super-orbital flux modulation is detected at $>3\sigma$ at the beginning and end of the lightcurve, increasing in period from $\sim$38 days to $\sim$44 days. Middle - The number of \swiftxrt\ observations used in each time bin. Right - The average \swiftxrt\ count rate observed in each time bin.}
\label{fig_epfold_lstat2_ulx7}
\end{figure*}

In \cite{brightman20}, we used simulations to determine the false alarm rate for the 38-day period signal, finding it to be $>99.9$\% significant. We repeat this analysis for our 300-day bins, simulating 10,000 lightcurves with 2000\,s resolution and a red noise power spectrum and we sample them with the same observational sampling as the real lightcurves. We then note the largest peak in each periodogram, irrespective of period. We define the false alarm rate as the number of simulated lightcurves that produce a peak as high as the real one, divided by the total number of simulations. We do this for each of the time bins since each time bin has a different number and spacing of observations. For each time bin, we determine the $L$-stat level which corresponds to a 0.3\% false alarm rate, equivalent to a 3$\sigma$ detection. In Figure \ref{fig_epfold_lstat2_ulx7} we plot the super-orbital periods which are detected at $>3\sigma$ against time.

In order to determine the uncertainties on the periods detected above, we use a Monte-Carlo technique to resample the lightcurve where the count rate of each observation is a random number that is taken from a Gaussian distribution with a mean equal to the observed count rate and a standard deviation equal to the uncertainty on the count rate. We do this 10,000 times, rerunning our analysis on each lightcurve, and noting the peak period. We define the 1$\sigma$ uncertainty as the standard deviation of periods recovered. We plot these derived uncertainties in Figure \ref{fig_epfold_lstat2_ulx7}.

We find that at the beginning of the lightcurve the super-orbital period is detected at $>3\sigma$ at $\sim$38 days, as found in \cite{brightman20}, but at the end of the lightcurve it is detected at $\sim$44 days. The highest peaks in the periodograms are at 0--300 days, with $P=38.2\pm0.5$ days, and 780--1080 days where $P=44.2\pm0.9$ days. These time bins are independent and the findings indicate that the super-orbital period has increased in length. There are no obvious correlations between the super-orbital period and the number of observations per bin, or the average count rate.

While the above approach accounts for the statistical uncertainties on the individual measurements, it does not account for possible effects due to sampling. In order to do this we employ a bootstrapping method, whereby we randomly exclude 10\% of the observations in each time bin. Repeating this 10,000 times, we find standard deviations on the periods derived were 0.13 for the 38.2 day period 0.58 for the 44.2 day period. If we increase the number of excluded pointings to 20\%, this becomes 0.28 and 1.83 respectively. This implies the statistical uncertainty on the measurements dominates the uncertainty in the super-orbital period.

Finally, we also check if the longer super-orbital period seen at 780--1080 days can be produced by the 38.2-day signal, but with the lightcurve sampling at 780--1080 days. We do this by projecting the 38.2 day profile to 780--1080 days and simulating a lightcurve with the same observing strategy during that time. Again, we do this 10,000 times, however none of these simulations produce a peak at 44.2-days or greater with a height at least equal to the observed periodogram. It is therefore highly unlikely that the \swift\ observing strategy caused the observed change in super-orbital period.

\begin{figure}
\begin{center}
\includegraphics[trim=20 0 20 20, width=85mm]{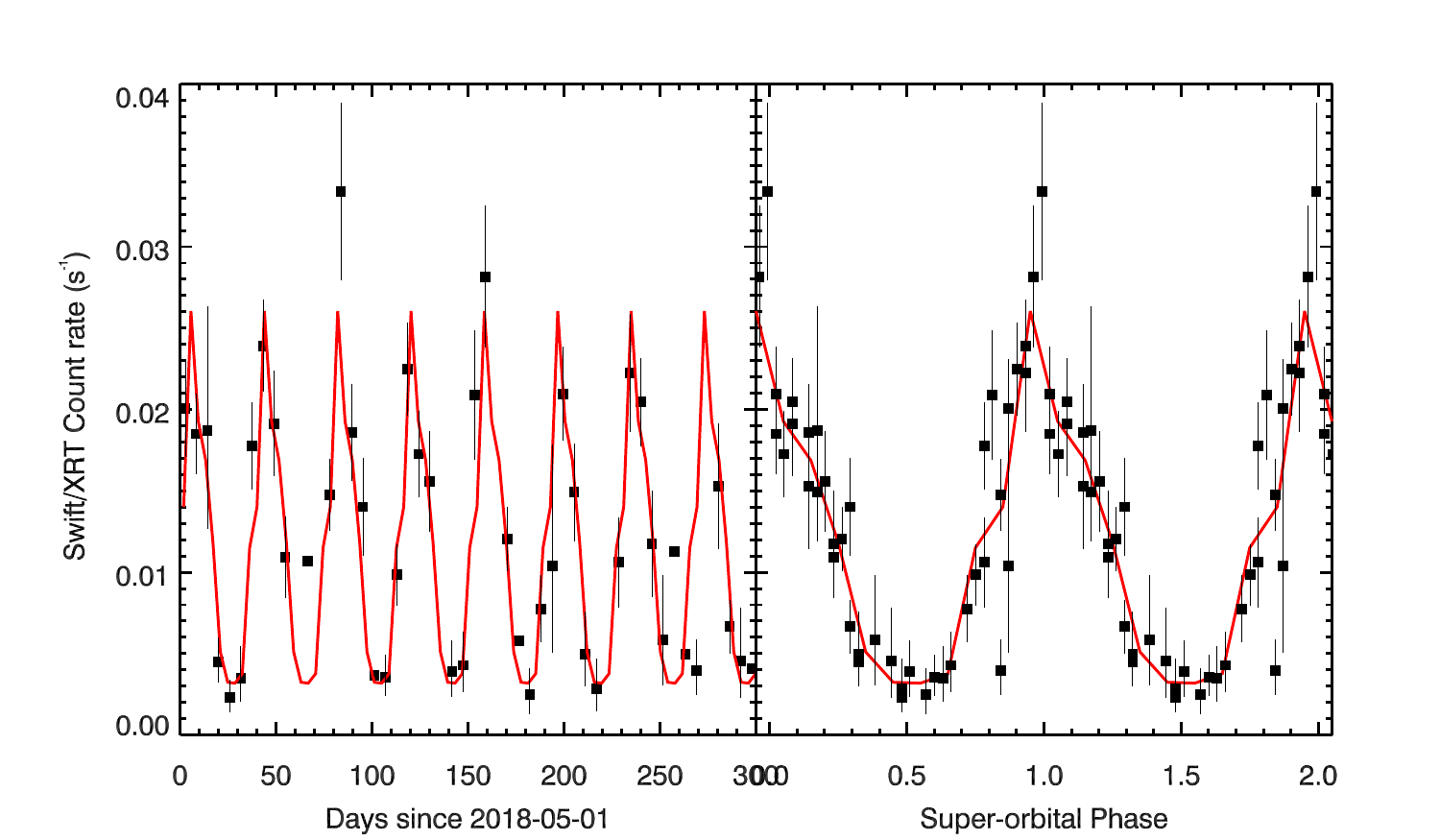}
\includegraphics[trim=20 0 20 20, width=85mm]{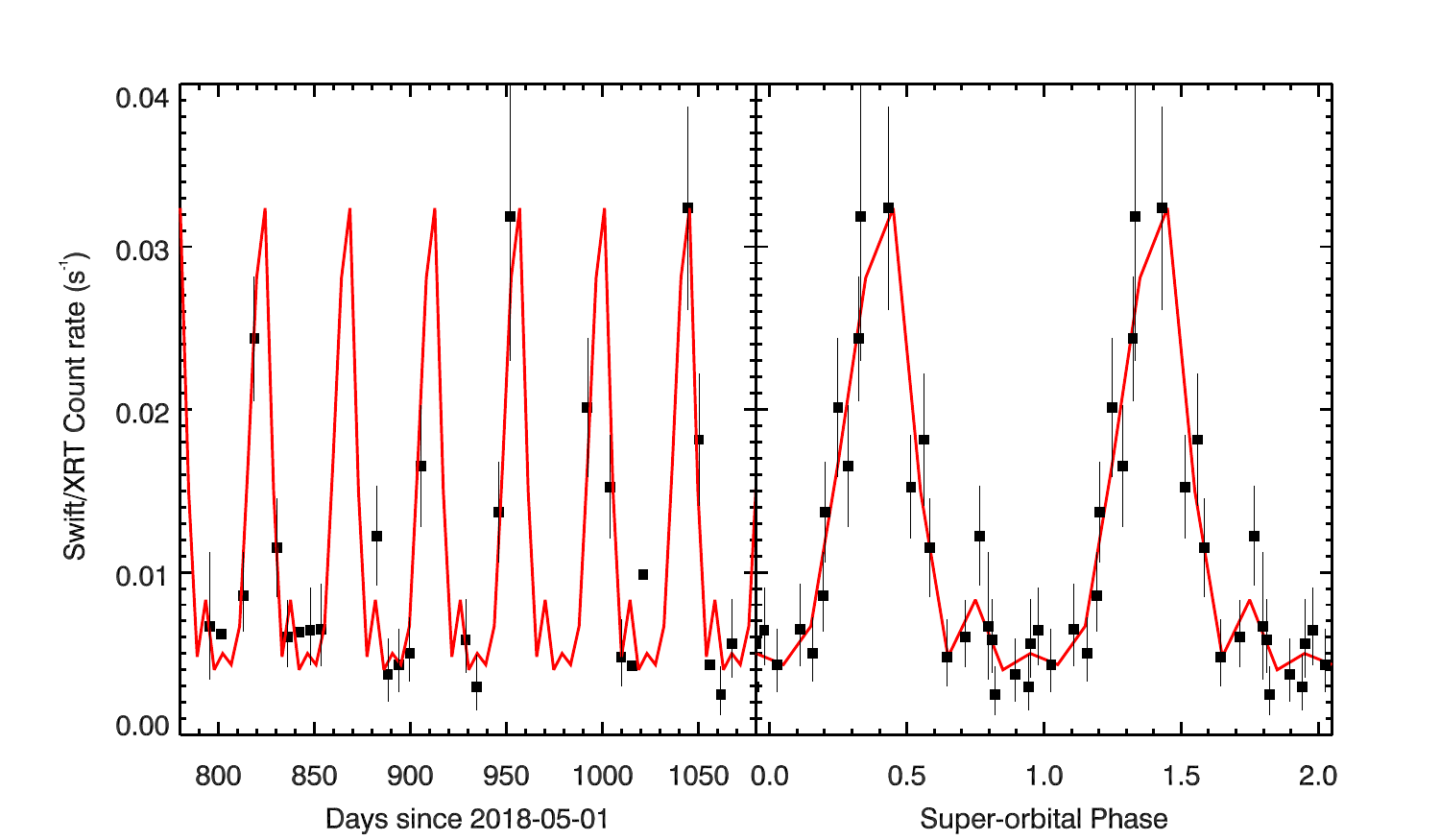}
\end{center}
\caption{\swiftxrt\ lightcurves and folded lightcurves over the 0--300 days and 780--1080 days epochs, showing the change in super-orbital period.}
\label{fig_epfold_lstat_profile}
\end{figure}

We show the lightcurves and folded lightcurves from the 0--300 days and 780--1080 days epochs in Figure \ref{fig_epfold_lstat_profile}. The profiles from the two lightcurves appear similar, albeit perhaps with most recent one being more peaked, and less sinusoidal.

\begin{table*}
\centering
\caption{Details of the observations used in this work}
\label{tab_obs}
\begin{center}
\begin{tabular}{c c c c c c c}
\hline
Observatory & ObsID 		& Start time 			& Exposure$^a$	& Count rate 	& Flux$^b$ & Super-orbital phase \\
 		& 				&	  (UT)			& (ks)	& (\cntrt)	& (\ergcms) & \\
\xmm\ 	& 0824450901		& 2018-05-13 21:18:47	& 60.6	& 0.167$\pm0.002$		& 5.8$\times10^{-13}$ & 0.15 \\
\xmm\ 	& 0830191401		& 2018-05-25 20:26:58	& 77.6	& 0.018$\pm0.001$		& 2.9$\times10^{-14}$ & -0.53 \\
\xmm\ 	& 0830191501		& 2018-06-13 01:39:03	& 48.2	& 0.221$\pm0.002$		& 7.9$\times10^{-13}$ & -0.04 \\
\xmm\ 	& 0830191601		& 2018-06-15 01:24:21	& 49.3	& 0.209$\pm0.002$		& 7.5$\times10^{-13}$ & 0.01 \\ 
\nustar\	& 60501023002	& 2019-07-10 05:56:09 	& 162	& 0.006$\pm0.0002$		& 7.6$\times10^{-13}$ & 0.26 \\
\xmm\ 	& 0852030101		& 2019-07-11 10:47:26	& 58.8	& 0.209$\pm0.002$		& 7.6$\times10^{-13}$ & 0.26 \\
\hline
\end{tabular}
\tablecomments{$^a$ After filtering  $^b$ 0.3--10 keV, observed (absorbed)}
\end{center}
\end{table*}

\section{Discussion}
\label{sec_disc}

\subsection{The spin of the neutron star}

The pulsations detected with $P=2.78674$\,s in our 2019 \xmm\ dataset of M51 ULX7 represent a secular spin up of the neutron star from $P=2.79812$\,s in 2018 \citep{rodriguez20}, with an average spin up rate of 3$\times10^{-10}$ s\,s$^{-1}$ over the 425-days spanning the observations. As described in Section \ref{sec_timing}, we were not able to correct the 2019 data for the orbital motion of the pulsar. We determined that this produced an uncertainty of order 1\,ms in the period. This then leads to an uncertainty of 0.5$\times10^{-10}$ s\,s$^{-1}$ in the derived secular spin-up rate. The secular spin up rate is similar to that calculated for the short timescale covered by the 2018 data, and appears comparable with the spin up of $\sim10^{-9}$ s\,s$^{-1}$ calculated from observations spanning 13 years, which is in the range of other ULX pulsars \citep{rodriguez20} .

\subsection{The broadband X-ray spectrum}

We presented here the first high-quality broadband X-ray spectrum with coverage above 10 keV of M51 ULX7, obtained with \nustar\ and \xmm, over the energy range 0.2--20 keV. Other members of the known ULX pulsar population with at least one high-quality broadband X-ray spectrum are NGC~5907~ULX \citep{walton15b}, NGC~7793~P13 \citep{walton18a}, and NGC~300~ULX \citep{carpano18}. M82 X-2 has broadband spectra but with significant confusion due to the proximity of M82 X-1 \citep{brightman20a}, and NGC 1313 X-2 is not well detected above 10 keV \citep{bachetti13}. Therefore adding to this small sample of ULX pulsars with a high-quality broadband X-ray spectrum is significant. Indeed, the sample of ULXs with high quality broadband spectra in general, be it with a known neutron star accretor or an unknown accretor, is also small \citep{walton18a}.

The broadband X-ray spectrum of M51 ULX7 can be well described by two disk-like components, plus a higher energy component associated with the pulsations. This is qualitatively very similar to the other ULX pulsars, and indeed, as pointed out in \cite{walton18c}, very similar to all ULXs in general, regardless of the known or unknown accretor \citep[see also][]{pintore17,koliopanos17}. As presented by \cite{walton18c}, the temperature of the cool component for ULXs ranges from 0.2--0.5 keV. We find for ULX7 that it is 0.33 keV. The hotter component has a temperature range of 1.2--3 keV, and is 2.5 keV for ULX7, making ULX7 indistinguishable from other ULXs in terms of its disk temperatures. \cite{walton18c} also noted that the temperature ratio for ULX pulsars in their sample was $\sim3$, while the other ULXs had a temperature ratio of $\sim8$. For the ULX7, the ratio is 7.5, therefore more in line with the other ULXs, rather than the other ULX pulsars. \cite{walton18c} also found that the flux ratio of the pulsed component, modeled by {\tt cutoffpl} to the total flux in the 0.3--40 keV band was higher for ULX pulsars than for the other ULXs. For M51 ULX7, however, this ratio is 0.15, which is relatively low in comparison to NGC~5907~ULX (0.82 in the high state), and NGC~7793~P13 (0.59).

\subsection{The spectral evolution}

The spectral evolution of ULX7 has been studied before. Both \cite{yoshida10} and \cite{earnshaw16} fitted the available \chandra\ and \xmm\ data at the time with a power-law model. \cite{earnshaw16} found that the spectral slope did not change significantly, despite the large changes in \lx. This included the very faint states observed with \chandra. \cite{rodriguez20} also investigated the evolution of the spectral parameters in their high-quality \xmm\ data from 2018, using their disk+black body model, also finding limited evidence for spectral variations, as did \cite{gurpide21}.

In our investigation of the spectral parameters of ULX7 and how they depend on the super-orbital phase of the system, we have found that the flux modulations are primarily driven by changes in the flux of the hotter disk-like component, with the flux of the cooler disk component changing less dramatically. We also see a potential dependence of both the disk temperature and normalization of the cooler disk-like component on super-orbital phase, even though the flux of this component does not change significantly with the super-orbital phase. However we do not have enough data to claim a significant relationship.

Since the normalization of the {\tt diskbb} and {\tt diskpbb} components are directly related to the inclination of the disk (also the distance to the source and inner disk radius), this potential dependence could be straightforwardly interpreted as disk precession, which has been suggested as the mechanism for the super-orbital periodic flux variations seen in ULX pulsars \citep[e.g.][]{fuerst17,dauser17,middleton18}. Indeed, \cite{fuerst17} also found evidence that the spectral parameters of the disk-like component of NGC~5907~ULX were dependent on the super-orbital phase of the system. In this case it was the hot disk-like component, rather than the cool one we see it in, and it was the radial temperature index, $p$, rather than the normalization/temperature. Since NGC~5907~ULX is relatively absorbed, the cool disk component is not visible. We note, however, that the models used are purely phenomenological, and any physical interpretations should have this added caveat.

The lack of any strong evolution of \nh\ with super-orbital period rules out a warped accretion disk which periodically obscures the X-ray source. While small variations in \nh\ can be seen, much larger changes are needed to produce the $>1$ magnitude flux modulation.

\subsection{The evolution of the super-orbital period}

Having discovered that the X-ray flux from M51 ULX7 is modulated on a period of 38.2$\pm0.5$ days in \cite{brightman20} from 2018--2019, we find here that period is variable, and has shifted to 44.2$\pm0.9$ days in 2020--2021. \cite{vasilopoulos20} also explored the \swiftxrt\ data on ULX7, finding the same 38-day period and also found tentative evidence for a 49 day period from \swiftxrt\ data taken in 2011, albeit covering only 1--2 cycles and therefore quite uncertain. 

An example of a super-orbital flux modulation in a source where the period is variable is SMC X-1, which has a super-orbital period of $\sim60$ days \citep{gruber84}, and exhibits recurrent excursions to shorter super-orbital periods, which may be (quasi)periodic themselves \citep{hu19}. Unlike M51 ULX7, SMC X-1 shows strong spectral variability across its super-orbital phase, with a harder spectrum at low fluxes, suggestive of absorption being the main cause. This absorption appears weaker during the super-orbital period excursion. From modeling of the pulse profile with super-orbital phase from \xmm\ and \nustar\ data, \cite{brumback20} found that the pulse shape and phase of SMC X-1 are consistent with reprocessed emission from a precessing inner disk. The wind-fed HMXB IGR J16493-4348 also exhibits a variable super-orbital period, however in that case, the amplitude is found to be variable, rather than the period. They suggest this is linked to a variable accretion rate. However, they note that the timing and spectral properties of ULXs show significant differences compared to those observed in wind-fed HMXBs such as IGR J16493-4348.

As discussed above, the evolution of the spectral parameters across super-orbital phase support the hypothesis that the flux modulations from M51 ULX7 are caused by precession of the accretion disk. Other theories such as a warped accretion disk periodically obscuring the X-ray source as suggested for SMC X-1 can be ruled out by the lack of \nh\ variations. It has also been suggested that a third orbiting star could be causing the variations \citep[e.g.][]{middleton18,rodriguez20}. This now appears unlikely since if the orbit of the third star were causing the periodic variations, the orbit would have needed to change significantly in a short time.

\cite{vasilopoulos20} discussed the potential for free-precession of the neutron star, which has been invoked to explain long term periodic changes of isolated neutron stars. However, they note that free precession of the neutron star alone cannot account for the variation in super-orbital period, which we find here.

\section{Conclusions}

In our new \nustar\ and \xmm\ data on M51 ULX7, we have found that the neutron star powering the source has spun up at a rate of  3$\pm0.5\times10^{-10}$ s\,s$^{-1}$ since the previous observations by \xmm, which is similar to that seen in other ULX pulsars. The data also provide the first high-quality broadband spectrum, consisting of two disk-like components, the temperatures of which are indistinguishable from other ULXs, and a high energy tail. We found that the luminosity of the hotter component drives the super-orbital flux modulation seen from the source. Finally we discovered that the super-orbital period varies, and has increased from 38.2$\pm0.5$ days in 2018--2019 to 44.2$\pm0.9$ days in 2020--2021. This change in period rules out some alternative explanations of the super-orbital period. 

\facilities{\swift (XRT), \nustar, \xmm} 

\software{{\tt NuSTARDAS}, {\tt XMMSAS} \citep{gabriel04}, {\tt XSPEC} \citep{arnaud96}, {\tt hendrics} \citep{bachetti15}}

\acknowledgements{

We thank the referee for their thorough review of our paper which improved it. 

This work was also supported under NASA Contract No. NNG08FD60C. \nustar\ is a project led by the California Institute of Technology, managed by the Jet Propulsion Laboratory, and funded by the National Aeronautics and Space Administration. This research has made use of the NuSTAR Data Analysis Software (NuSTARDAS) jointly developed by the ASI Science Data Center (ASDC, Italy) and the California Institute of Technology (USA).

This work was also based on observations obtained with XMM-Newton, an ESA science mission with instruments and contributions directly funded by ESA Member States and NASA

We wish to thank the \swift\ PI, Brad Cenko for approving the target of opportunity requests we made to observe M51, as well as the rest of the \swift\ team for carrying them out. We also acknowledge the use of public data from the \swift\ data archive. This work made use of data supplied by the UK Swift Science Data Centre at the University of Leicester. 

This research has made use of data and/or software provided by the High Energy Astrophysics Science Archive Research Center (HEASARC), which is a service of the Astrophysics Science Division at NASA/GSFC.

GLI  acknowledges  funding  from  the  Italian  MIUR PRIN  grant  2017LJ39LM. The work of DS was carried out at the Jet Propulsion Laboratory, California Institute of Technology, under a contract with NASA. DJW acknowledges support from an STFC Ernest Rutherford Fellowship.

}

\bibliography{manuscript.bbl}

\end{document}